\documentstyle[12pt,a4,epsfig]{article}
\newcommand{\bmath}{\begin{displaymath}}
\newcommand{\emath}{\end{displaymath}}

\title {
\begin{flushright}
{\normalsize TAUP-2244-95\\ WIS-95/15/Mar-PH}
\end{flushright}
\vspace{2.0cm}
Spin Contents of Nucleons with SU(3) Breaking}
\author {Jechiel LICHTENSTADT$^1$\thanks{Supported by the Basic Research Fund
of the Israeli Academy of Sciences} and Harry J.
LIPKIN$^{1,2}$\thanks{Supported
in part by grant No. I-0304-120-.07/93 from The German-Israeli Foundation
for Scientific Research and Development} \\
$^1$School of Physics and Astronomy\\
The Raymond and Beverly Sackler Faculty of Exact Sciences\\
Tel Aviv University, 69978 Tel Aviv, Israel\\
and\\
$^2$Department of Nuclear Physics\\
The Weizmann Institute of Science, 76100 Rehovot, Israel\\
}
\date { }
\begin {document}
\maketitle
\begin{abstract}

    We apply a model for SU(3) breaking to the analysis of
the spin contents of the nucleon from the the latest data on first
moments of the spin dependent structure functions and include higher
order QCD corrections. The results show that the value of the total
quark spin contribution to the nucleon spin
$\Delta \Sigma \equiv \Delta u + \Delta d +\Delta s$
remains about 0.3 and is very insensitive to SU(3) breaking, while the
result for the strange quark contribution varies considerably with SU(3)
breaking.
\end{abstract}
%
%
\newpage

     Recent experiments of polarized deep inelastic scattering (DIS) provided
us
 with
high quality data for the spin structure functions of the proton, deuteron and
 neutron~\cite{SMC93a,E142,SMC94b,E143}.
These measurements are used to evaluate the first moments of the spin dependent
structure functions which can be interpreted in terms of the contributions of
 the
quark spins ($\Delta \Sigma =\Delta u + \Delta d +\Delta s$) to the total spin
of the nucleon. The first results from the measurements of the proton spin
 structure function
by the EM Collaboration~\cite{As88} were very surprising, implying that $\Delta
 \Sigma$ is
 rather small
(about 10\%) and that the strange $sea$ is strongly polarized. The new
 measurements~\cite{SMC94b,E143},
together with the new values of $F$ and $D$~\cite{Cl93,Hs88}, suggest that
 $\Delta \Sigma$
is significantly larger than what was inferred from the EMC experiment,
 concluding that
$\Delta \Sigma \approx 0.27 \pm 0.14$ and $\Delta s = -0.10 \pm 0.05$.
Incorporating also the recent SLAC E143 data for the proton with all other
proton data, has not changed these results, and the combined analysis
yields~\cite{SMC95}  for the first moment $\Gamma_1^p(Q_0^2=5GeV^2) = 0.137 \pm
 0.010$
and $\Delta \Sigma = 0.28 \pm 0.09$ and $\Delta s = -0.10 \pm 0.03$

    With $\int_0^1 g(x)dx = {1 \over 2}\sum_{flavors} e_f^2 \Delta q_f$
the underlying hypothesis in these analyses is the quark parton model (QPM)
 interpretation
of the neutron weak decay constant $g_A = \Delta u - \Delta d$ and the SU(3)
 flavor
symmetry interpretation of $3F-D = \Delta u + \Delta d - 2\Delta s$

    It is usually implied that the experimental evaluation of $\Delta \Sigma$
 and $\Delta s$
is based on exact  SU(3) flavor symmetry. However, the extracted values
 are
always quoted with errors which are derived from the experimental errors
on the measured quantities ($\Gamma_1, g_A$ and $3F-D$). Little attention has
 been given
to evaluating the effect of SU(3) breaking on these results.

We first examine the information obtainable from experimental data to
separate results which do not assume SU(3) from those which do.
We will see that the value of $\Delta \Sigma$ deduced from the measured
first moment of the spin dependent structure function is not sensitive
to the assumption of exact SU(3), or to the values of the SU(3)
parameters $F$ and $D$.

   In the quark parton model, $\Gamma_1^p$ can be expressed as

\begin{equation}
 \Gamma_1^{p} = \int_0^1 g_1^{p}(x) dx =C_{ns}({\alpha_s \over \pi}){1
 \over 12}
\left[
{4\Delta u -2\Delta d - 2\Delta s
\over 3} \right]
+ C_s({\alpha_s \over {\pi} }) {1 \over 9}
\Delta \Sigma
\end{equation}
where $C_{ns}({\alpha_s \over \pi})$ and $C_s({\alpha_s \over \pi})$
include the higher order QCD corrections which are taken from
Ref.~\cite{Larin94} for the linear combinations of the matrix elements
of the axial-vector current which transform under SU(3)
like members of an octet and singlet respectively.
Here, anomalous contributions to $\Delta \Sigma$ in the singlet term
were neglected, and $\Delta u, \Delta d$ and $\Delta s$ are assumed to be
the same in the singlet and non-singlet term.
At this stage there
is no assumption of SU(3) symmetry for the nucleon wave function. There
is therefore no relation between the values of the contributions to the
quark spins $\Delta u$, $\Delta d$ and $\Delta s$ and other parameters
determined from weak decays.

When SU(3) symmetry is assumed for the octet baryon wave function, the
expression ${4\Delta u -2\Delta d - 2\Delta s}$ can be expressed entirely
in terms of the quantities measured in weak decays ($g_A$ and $3F-D$).
Substituting

\begin{equation}
3F-D = \Delta u + \Delta d - 2\Delta s =
\Delta \Sigma -3\Delta s
\end{equation}
gives the conventional expression,
\begin{equation}
 \Gamma_1^{p(n)} = \int_0^1 g_1^{p(n)}(x) dx =C_{ns}({\alpha_s
\over \pi}){1  \over 12}
\left[+(-) g_A + {3F-D \over 3} \right]
+ C_s({\alpha_s \over {\pi} }) {1 \over 9}
\Delta \Sigma
\end{equation}
(where the (-) refers to the neutron case).
A similar expression can be written for the
deuteron first moment which can be expressed as a sum of those of the
proton and neutron:
\begin{equation}
\Gamma_1^d = (\Gamma_1^p + \Gamma_1^n) ({ 1 - 1.5\omega_D \over 2 })
\end{equation}
where corrections for the D-state contributions in the deuteron ground
state were considered.

It is interesting to examine the case
when only isospin symmetry is assumed for the nucleon wave function. The
expression ${4\Delta u -2\Delta d - 2\Delta s}$ can be written in terms
of the quantity $g_A$ measured with high precision in the neutron decay,
$\Delta \Sigma$ and $\Delta s$:
\begin{equation}
 \Gamma_1^{p(n)} = \int_0^1 g_1^{p(n)}(x) dx =C_{ns}({\alpha_s
\over \pi}){1  \over 12} \left[+(-) g_A + {\Delta \Sigma\over 3}
-\Delta s \right]
+ C_s({\alpha_s \over {\pi} }) {1 \over 9}
\Delta \Sigma
\end{equation}

\begin{figure}[h]
\epsfig{figure=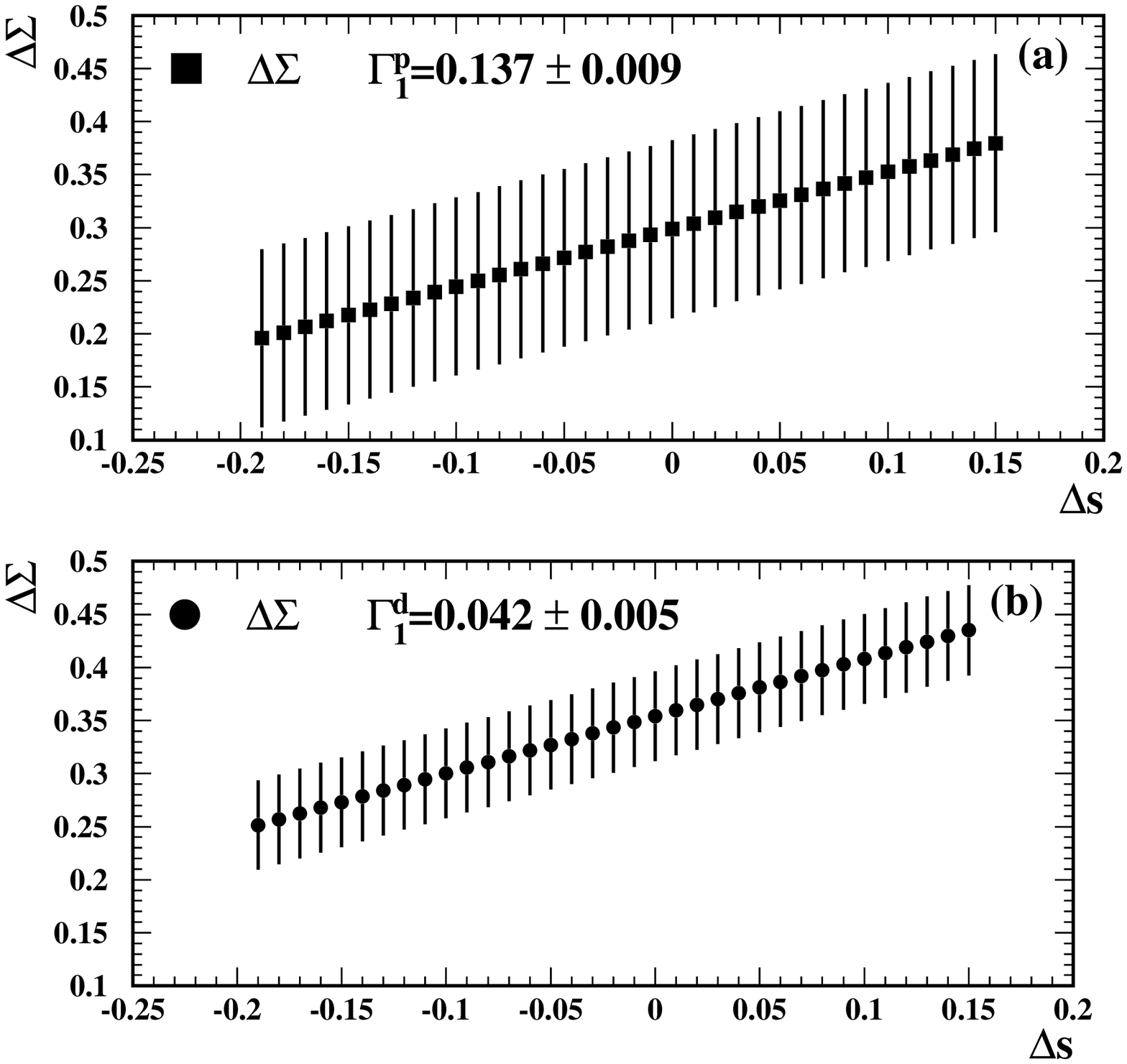,width=14cm,height=14cm}
\caption{$\Delta \Sigma$ vs $\Delta s$ for the proton (a)
and for the deuteron (b).
 }
\label{dsigds}
\end{figure}

We can then
express $\Delta \Sigma$ as a function of $\Delta s$ for given values
of $g_A$ and $\Gamma_1^N$.
Taking the result for the ``world" value for the proton~\cite{SMC95} :
$\Gamma_1^p (Q_0^2$ = 5 GeV$^2$) = 0.137 $\pm$ 0.010  and the recent
results of SLAC E143 for the deuteron~\cite{E143a}
$\Gamma_1^d (Q_0^2$ = 3 GeV$^2$) = 0.042 $\pm$ 0.005
we show in Fig.~\ref{dsigds} the dependence of $\Delta \Sigma$ with the
 variation
of $\Delta s$ from -0.2 to +0.15. We can conclude
that $\Delta \Sigma$ obtained from the spin structure
function first moment is well established, and variations
of $\Delta s$ by some model (like using input from hyperon decays with
SU(3) breaking) will have a relatively small effect on $\Delta \Sigma$.

Introducing SU(3) breaking into the analysis of hyperon decays is
nontrivial and model dependent. The quantity denoted by $g_A$ is really a
ratio of axial-vector and vector matrix elements. Both matrix elements
can be changed by SU(3) symmetry-breaking, but it is only the axial
matrix element that is relevant to the spin structure. The information
from hyperon decays used in conventional treatments of spin structure
is expressed in terms of $D$ and $F$ parameters which characterize the
axial couplings only under the assumption that the vector coupling is
pure $F$ and normalized by the conserved vector current,
where the whole SU(3) octet currents are conserved.
 Thus any attempts to parameterize SU(3) breaking in fitting
hyperon data by defining ``effective" $D$ and $F$ parameters immediately
encounter the difficulty of how much of the breaking comes from the
axial couplings and how much comes from the vector
and the breakdown of the conserved vector currents for
strangeness changing currents. The vector matrix
element is uniquely determined by Cabibbo theory in the SU(3) symmetry
limit. The known agreement of experimental vector matrix elements
with Cabibbo theory places serious constraints on possible SU(3) breaking
in the baryon wave functions. On the other hand the strange quark
contribution to the proton sea is already known from experiment
be reduced roughly by a factor of two from that of a flavor-symmetric
sea~\cite{CCFR}.

To assess the the uncertainty on $\Delta \Sigma$ and $\Delta s$ due to
flavor symmetry breaking we use a model consistent with the above
constraints~\cite{HJL94} suggested by one of us (H.~J.~L.),
to incorporate the weak decay data with the polarized  DIS results.
The good results of Cabibbo theory for the weak vector decays are kept
by assuming that the weak current strangeness changing
matrix elements are due to a strangeness change in a valence quark,
while the sea wave functions have a nearly 100\% overlap and
that the valence quark wave functions satisfy SU(3).
In this picture it is convenient to express $\Delta q$ as the sum of its
valence and sea contributions, $\Delta q = \Delta q^v + \Delta q^s$.
We immediately obtain
the result that the SU(3) fits to hyperon decay data give information
only about the wave functions of the valence quarks and the values of
the $D$ and $F$ parameters give information about the spin contributions
$\Delta u^v$, $\Delta d^v$ and $\Delta s^v$ from the valence quarks.

In this model the flavor SU(3) symmetry breaking relevant for the
$\Sigma^- \rightarrow n$ decay is realized by a suppression of the
strange pair production in both neutron and $\Sigma^-$ seas, and
described by a parameter $\epsilon$. Noting that the neutron has $no$
valence strange quarks, and that $\Delta u^s $ is the $same$ for both
neutron and $\Sigma^-$ seas the following results were
obtained~\cite{HJL94}
\begin{equation}
\Delta s^v(n) = 0; ~ ~ ~ ~
 \Delta u^s(\Sigma^-) \approx \Delta u^s(n) = (1+\epsilon)\Delta s(n)
\end{equation}

\begin{eqnarray}
& & \ \ \ \ \ \ \ {G_A \over G_V}(\Sigma^- \rightarrow n)  =  F-D  =  \Delta
 u^v(n) -\Delta
 s^v(n)=  \nonumber \\
& &  = \Delta u(n) - \Delta s(n) - \left[ \Delta u^s(n) -   \Delta  s^s(n)
 \right]
= \Delta u(n) - (1+\epsilon)\Delta s(n)
\end{eqnarray}

Combining this result with the corresponding relation for the neutron
weak decay constant (and SU(2) isospin symmetry):

\begin{equation}
 g_A = F+D = \Delta u(p) -\Delta d(p)= \Delta d(n) - \Delta u(n) = 1.257 \pm
 0.003
\end{equation}
we obtain  for the proton or neutron:
\begin{equation}
3F-D = 2\left[ {G_A \over G_V }(\Sigma^- ) \right] + g_A = \Delta u(N) +
 \Delta d(N)
- 2(1+\epsilon)\Delta s(N) = 0.575 \pm 0.016
\end{equation}
or
\begin{equation}
 3F-D = {3 + 2\epsilon \over 3}(\Delta u + \Delta d - 2\Delta s) -
{2\epsilon \over 3}\Delta \Sigma
\end{equation}
with the values for $g_A$ and $3F-D$ taken from~\cite{Cl93,Hs88}.

\begin{figure}[h]
\epsfig{figure=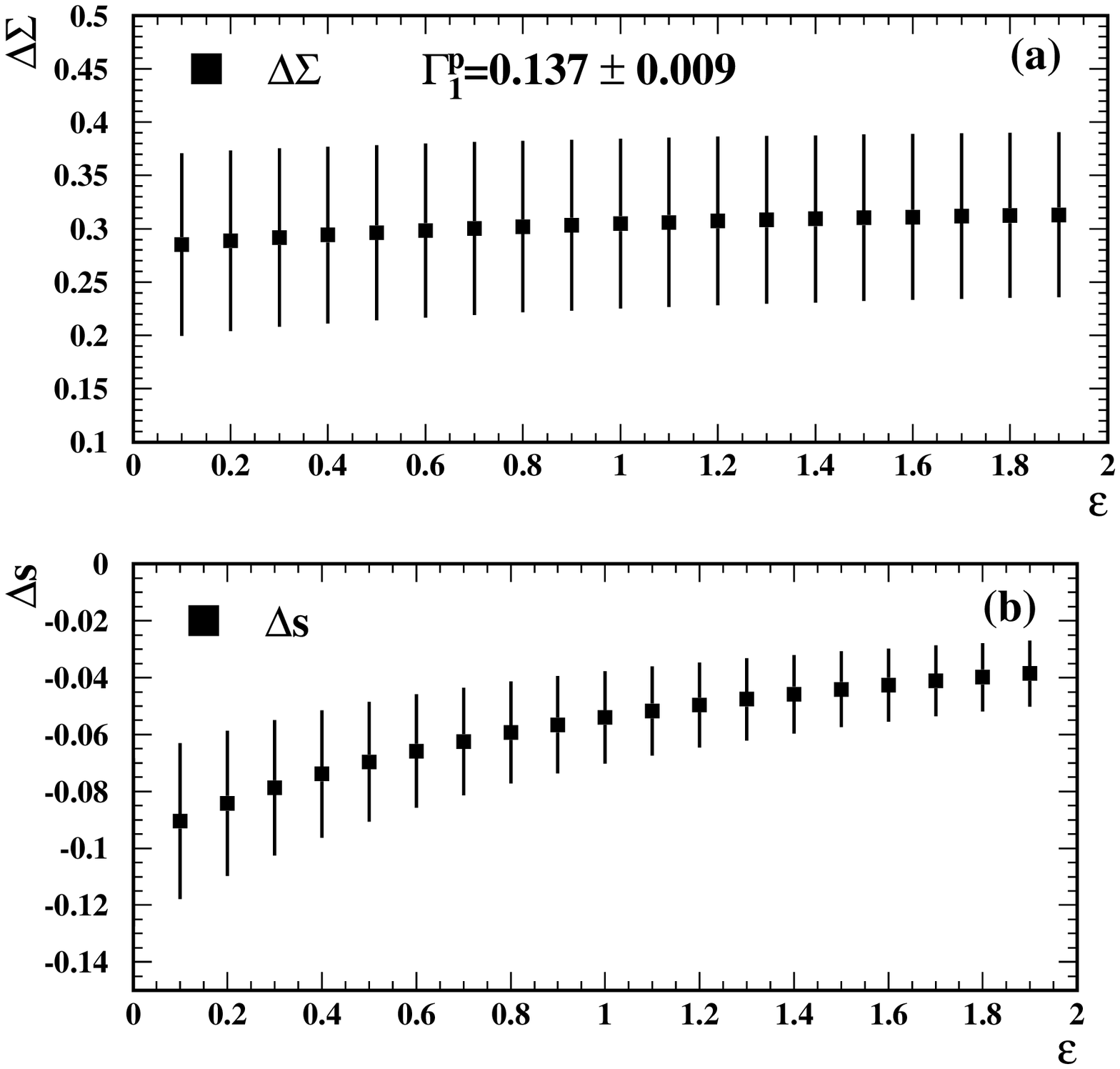,width=14cm,height=14cm}
\caption{$\Delta \Sigma$ (a) and  $\Delta s$ (b) for the proton
as a function of the SU(3) symmetry breaking parameter $\epsilon$.
}
\label{su3p}
\end{figure}

\begin{figure}[h]
\epsfig{figure=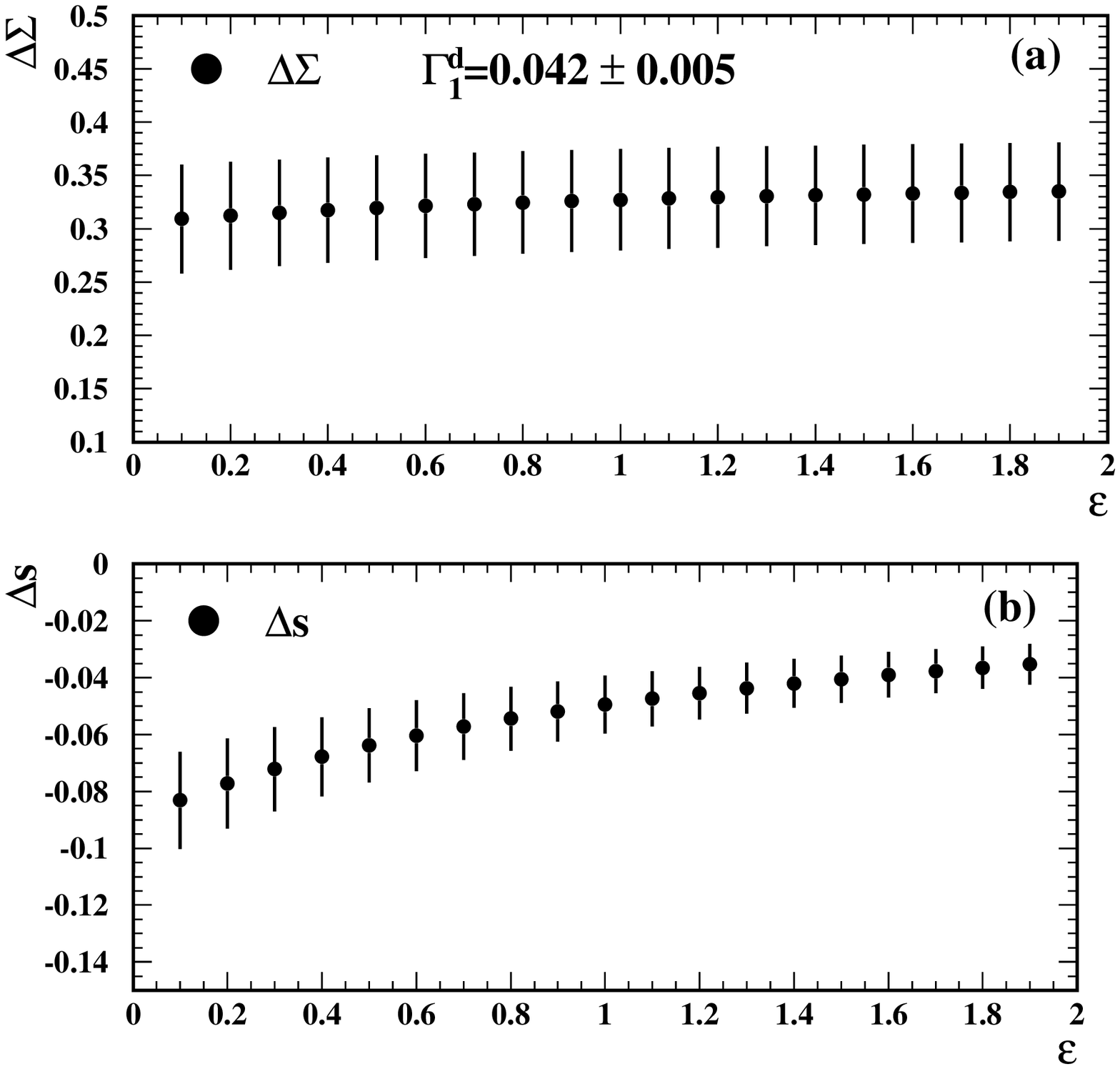,width=14cm,height=14cm}
\caption{$\Delta \Sigma$ (a) and  $\Delta s$ (b) for the deuteron
as a function of the SU(3) symmetry breaking parameter $\epsilon$.
}
\label{su3d}
\end{figure}

  With the new QPM interpretation of $3F-D$ under the flavor symmetry
breaking we can substitute $3F-D$ in Eq. 1 with
\begin{equation}
\Delta u + \Delta d - 2\Delta s = {3(3F-D) \over 3 +2\epsilon} +
{2\epsilon \over 3 + 2\epsilon} \Delta \Sigma
\end{equation}
which then reads:
\begin{equation}
 \Gamma_1^p = {C_{ns}({\alpha_s \over \pi}) \over 12}
\left[ g_A + {3F-D \over 3+2\epsilon} \right]
+ \left[ {C_s({\alpha_s \over \pi}) \over 9} + {C_{ns}({\alpha_s \over \pi})
 \over 12}
\cdot {2\epsilon \over 3(3+2\epsilon)} \right]
\Delta \Sigma
\end{equation}
and
\begin{equation}
 \Delta s = { \Delta \Sigma - (3F-D) \over 3 + 2\epsilon}.
\end{equation}
   Taking for experimental value for $\Gamma_1^p$, we  show in Fig.~\ref{su3p}
 the
 variations
of $\Delta \Sigma$ and $\Delta s$ as a function of the asymmetry parameter
 $\epsilon$.
There is hardly a noticeable change in $\Delta \Sigma$ which moves from 0.28
at $\epsilon = 0$ (which is the exact SU(3) symmetry limit), to 0.31 for
 $\epsilon = 2.$
On the other hand $\Delta s$ varies from -0.10 to -0.04, which implies no
significant contribution due to the strange sea if flavor symmetry is
significantly broken.
The same conclusions can be drawn from the neutron results or from the
deuteron measurement of $\Gamma_1^d$ as shown in Fig.~\ref{su3d}.

   In conclusion, the value of the quark spin contribution to the nucleon spin
 extracted
from the first moments of the spin dependent structure functions is relatively
independent of the assumption exact SU(3) flavor symmetry. The latter
affects the value obtained for the ``$strange~sea"$ contribution $\Delta s$.
It is interesting to assess
the anomalous contributions in the singlet term, and study the variations of
the
 extracted
spin contents with different models for the gluon contribution in the spin
 structure functions.

\newpage

\begin{thebibliography}{99}

\bibitem{SMC93a} SMC, B. Adeva et al., Phys. Lett. B302 (1993) 533.

\bibitem{E142}  E-142, P.L. Anthony et al.,
                Phys. Rev. Lett. 71 (1993) 959.
\bibitem{SMC94b} SMC, D. Adams et al.,
                 Phys. Lett. B329 (1994) 399.
\bibitem{E143}  E-143, K. Abe et al.,
                Phys. Rev. Lett. 74 (1995) 346.

\bibitem{As88} EMC, J. Ashman et  al.,
               Phys. Lett. B206, (1988) 364;
               Nucl. Phys. B328 (1989) 1.

\bibitem{Cl93}  F. E. Close and R. G. Roberts,
                 Phys. Lett. {\bf B 316} (1993) 165

\bibitem{Hs88}  S. Y. Hsueh  et al.,
                Phys. Rev.  D38, (1988) 2056.

\bibitem{SMC95} The SM Collaboration, to be published.

\bibitem{Larin94}  S.A. Larin,
                Phys. Lett. B 334 (1994) 192.

\bibitem{E143a}  E-143, K. Abe et al.,
                 SLAC-PUB-95-6734,
                 submitted to Phys. Rev. Lett.

\bibitem{CCFR} A. O. Bazarko et al (CCFR Collaboration), Nevis Preprint
R1502 (June 30, 1994), submitted to Physics Letters B

\bibitem{HJL94}  Harry J. Lipkin,
                 Phys. Lett.  B 337 (1994) 157.


%
\end{thebibliography}
\end {document}